# Franck-Condon Factors and Radiative Lifetime of the $A^2\Pi_{1/2}$ - $X^2\Sigma^+$ Transition of Ytterbium Monoflouride, YbF

Xiujuan Zhuang[a], Anh Le[a], Timothy C. Steimle[a], N. E. Bulleid[b], I. J. Smallman[b], R. J. Hendricks[b], S. M. Skoff[b], J. J. Hudson[b], B E. Sauer[b], E. A. Hinds[b] and M. R. Tarbutt[b]

[a]*Department of Chemistry and Biochemistry, Arizona State University, Tempe, Arizona 85287-1604, USA.*

[b]*Centre for Cold Matter, Blackett Laboratory, Imperial College London, Prince Consort Road, London SW7 2AZ, UK.*

**Abstract**

The fluorescence spectrum resulting from laser excitation of the $A^2\Pi_{1/2} \leftarrow X^2\Sigma^+$ (0,0) band of ytterbium monofluoride, YbF, has been recorded and analyzed to determine the Franck-Condon factors. The measured values are compared with those predicted from Rydberg-Klein-Rees (RKR) potential energy curves. From the fluorescence decay curve the radiative lifetime of the $A^2\Pi_{1/2}$ state is measured to be 28±2 ns, and the corresponding transition dipole moment is 4.39±0.16 D. The implications for laser cooling YbF are discussed.

**Introduction**

The recent demonstration of optical cycling and laser cooling in strontium monofluoride [1,2] suggests that other similar molecules may also be amenable to laser cooling [3]. YbF is a particularly interesting candidate because it is being used to measure the electric dipole moment of the electron [4,5]. The use of optical forces to produce slower and colder beams of YbF molecules would enable longer interaction times and hence an improved sensitivity in this measurement [6]. For laser cooling to be feasible, the state excited by the laser should have a short lifetime so that the scattering rate is high, and it should decay to only a small number of ro-vibronic levels to minimize the number of laser frequencies required [3]. The $A^2\Pi_{1/2} \leftarrow X^2\Sigma^+$ (0,0) transition of ytterbium monofluoride, YbF, is expected to meet these criteria. The transition is intense because, to a first approximation, it is a promotion of an electron from a non-bonding $6s/6p/5d$ σ-orbital to the non-bonding $6p_{\pm 1}/5d_{\pm 1}$ π-orbital both located on Yb$^+$. The associated minimal change in the intermolecular potential assures that the emission will be predominately back to the $X^2\Sigma^+$ ($v$=0) vibronic state.

The $A^2\Pi_{1/2} \leftarrow X^2\Sigma^+$ (0,0) band system of YbF has been extensively studied both at Doppler limited (≈1.4 GHz) resolution [7] and at sub-Doppler resolution using molecular beam techniques [8-14]. The Fourier transform $N$=1 ← $N$=0, $X^2\Sigma^+$ ($v$=0 and 1) pure rotational microwave spectrum of the $^{174}$YbF isotopologue has also been recorded and analyzed [13]. An optimized set of field-free spectroscopic parameters for all six isotopologues ($^{170}$Yb(3.5%), $^{171}$Yb(14.3%), $^{172}$Yb(21.9%), $^{173}$Yb(16.1%), $^{174}$Yb (31.8%) and $^{176}$Yb (12.7%)) can be found in Ref. 13. The electric dipole moment, $\mu_e$(= 3.49 D), for the $X^2\Sigma^+$ ($v$=0) state was determined from the analysis of the Stark effect on the radio frequency transition between the $F$=1 and $F$=0 hyperfine levels of the $N$=0 rotational state [9]. The optical Stark effect in the $Q(0)$ line of the $A^2\Pi_{1/2}$ – $X^2\Sigma^+$ (0,0) band was analyzed to determine $\mu_e$(= 2.48D) for the $A^2\Pi_{1/2}(v$ =0) state [11]. The optical Zeeman spectra of the $A^2\Pi_{1/2}$ – $X^2\Sigma^+$(0,0) band of $^{171}$YbF, $^{172}$YbF and $^{174}$YbF isotopologues have recently been recorded and analyzed [15].

Here we report on the fluorescence spectrum resulting from excitation of the $R_1(2)(\nu=18109.454$ cm$^{-1})$ branch feature of the $A^2\Pi_{1/2} \leftarrow X^2\Sigma^+$ (0,0) band. Franck-Condon factors, $f_{v'-v''}$, are determined from the relative intensities under the assumption that the transition moment does not have a strong internuclear separation dependence in the region of the equilibrium bond distances. The spectroscopic parameters are used to predict the Franck-Condon factors and these are compared to the experimental values. The radiative lifetime is determined from the decay of fluorescence following pulsed excitation of lines in the $P_1+Q_{12}$ band head ($\nu=18106$ cm$^{-1}$).

### Experimental

The Franck-Condon factors were independently determined at both Imperial College (IC) and Arizona State University (ASU) using wavelength-resolved fluorescence detection.

The experimental setup used at IC, shown schematically in Figure 1, employs a pulsed supersonic beam of YbF seeded in helium carrier gas which is cooled to cryogenic temperatures. The new source is inspired by recent work on the formation of YbF in a cold buffer gas cell [16, 17]. The molecules are created by laser ablation of a Yb/AlF$_3$ target located inside an open-ended copper tube mounted on the low temperature cold plate of a closed-circuit cryocooler. A solenoid valve, thermally connected to the cold plate, is used to create a pulse of cold helium gas that passes through a thin tube into the region just below the target. The timing of the laser ablation pulse is such that YbF molecules are formed just as the gas pulse passes near the target. Collisions between the YbF molecules and the cold helium lead to cooling of the translational, rotational and vibrational modes of the molecules to approximately 4 K. The molecules entrained in the gas pulse form a well-directed beam that travels at about 300 m/s into the detection region. Although similar speeds and temperatures are obtained using room temperature xenon as the carrier gas [12], the present source provides far higher beam intensity

The molecules are illuminated with light from a cw single-mode ring dye laser, tuned to the $^{174}$YbF $R_1(2)(\nu=18109.454$ cm$^{-1})$ line of the $A^2\Pi_{1/2} \leftarrow X^2\Sigma^+$ (0,0) band. The resulting fluorescence is collected by an imaging system consisting of a spherical mirror and an aspheric lens placed on either side of the molecular beam. The collimated fluorescence is directed onto a plate beamsplitter which divides the fluorescence into "reference" and "signal" components. Each component is focused, spatially filtered and passed through an infrared blocking filter to eliminate 1064 nm light from the ablation laser, and is then detected with a photomultiplier tube (PMT). A 40 nm full-width half-maximum (FWHM) band pass filter centred on 550 nm is used to reduce ambient background light in the reference channel. Various bandpass and longpass filters are placed in the signal arm to study different spectral components of the fluorescence light. Bandpass filters centred at 550 nm, 570 nm, and 580 nm, each with ±10 nm FWHM, are used to measure the $A^2\Pi_{1/2}$ ($v'=0$) $\rightarrow$ $X^2\Sigma^+$ ($v''=0, 1$ and 2) fluorescence intensities. A longpass filter with nominal cut-off at 590 nm permits the measurement of any additional light due to transitions to higher vibrational states. The absolute transmissions of the beamsplitter, windows and filters are all separately measured at the 0-0, 0-1 and 0-2 wavelengths by measuring the laser power they each transmit. This, combined with the PMT's specified wavelength-dependent response, determines the relative sensitivity of the detection system at each wavelength. Each filter was used multiple times in a varying order and rotated at various angles about the normal, so as to randomise effects due to filter placement and to average away a small observed polarization dependence. The probe laser was modulated on and off on alternate ablation pulses to distinguish between the LIF and any background light. By using the ratio of the signal in the signal arm to that in the reference arm, the measurement is insensitive to variations in the probe laser power and the intensity of the molecular beam.

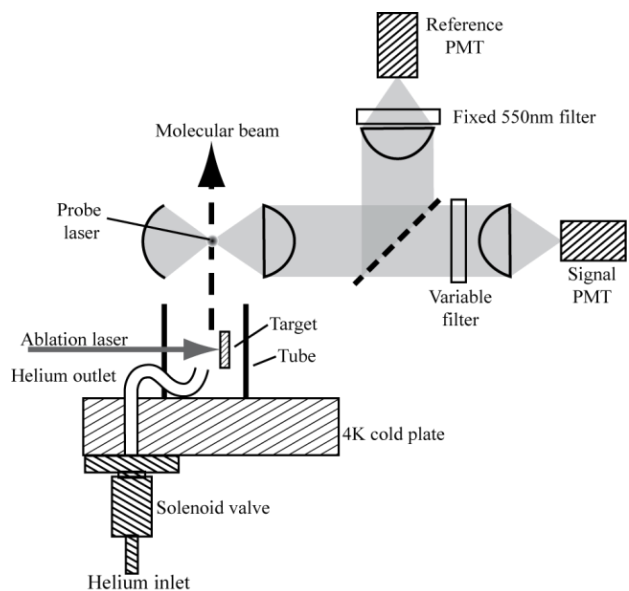

**Figure 1**: The apparatus used to create and detect cold YbF pulses at Imperial College.

The apparatus used at ASU for molecular beam production and detection of YbF has been previously described [14,15]. A molecular beam of YbF was generated by ablating a solid, rotating, ytterbium rod near a supersonically expanding mixture of 5% SF$_6$ and 95% argon. Excitation spectra were recorded using both a cw single-mode dye laser ($\Delta\nu < 1$ MHz) and a pulsed dye laser ($\Delta\nu \approx 3$ GHz) that provides 10 ns long pulses. In the high-resolution scans absolute frequencies were determined to an accuracy of ±3 MHz by simultaneously recording the sub-

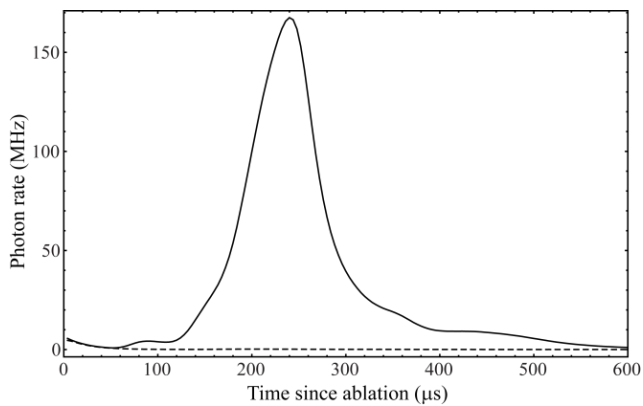

**Figure 2**: Signal from the reference PMT, with probe laser on (solid line) and off (dashed line), averaged over 100 shots. The initial decaying signal is light from the ablation process. The peak at approximately 250 µs is the LIF signal resulting from excitation of the $R_1(2)(\nu=18109.454\ \text{cm}^{-1})$ branch feature of the $A^2\Pi_{1/2} \leftarrow X^2\Sigma^+$ (0,0) transition of $^{174}$YbF.

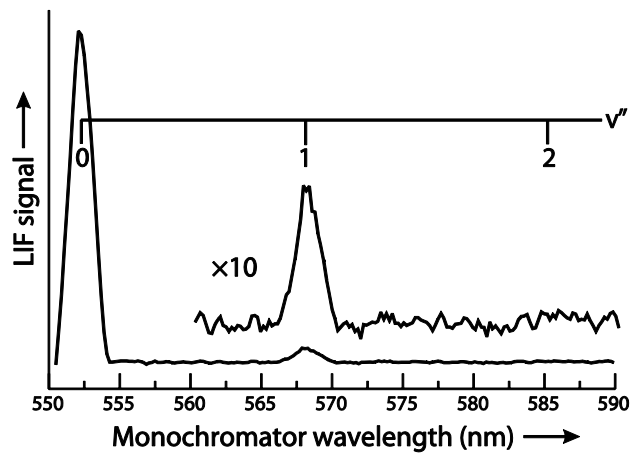

**Figure 3**: The DLIF spectrum resulting from excitation of the $R_1(2)(\nu=18109.454\ \text{cm}^{-1})$ branch feature of the $A^2\Pi_{1/2} \leftarrow X^2\Sigma^+$ (0,0) transition of $^{174}$YbF and viewed through a 2/3 m scanning monochromator.

Doppler $I_2$ spectrum. Between $I_2$ lines, frequency shifts could be measured to approximately 50 MHz by simultaneously recording the transmission of an actively stabilized étalon. Dispersed laser-induced fluorescence (DLIF) spectra were obtained by recording the LIF signal with a 2/3 meter scanning monochromator. The slits of the monochromator were adjusted to give a spectral resolution of ±1.5 nm. The digital delay controller set the relative timing of the laser ablation, pulse valve and gated photon counter. The wavelength dependence of the spectrometer sensitivity was determined by recording the emission of a calibrated tungsten filament. The fluorescence decay curves were recorded by replacing the single-mode ring dye laser with a pulsed dye laser and the gated photon counting system with a digital storage oscilloscope.

### Observations

Figure 2 shows two typical traces obtained from the IC experiment, one with the probe laser on and the other with it off, each an average of 100 molecular pulses measured by the reference PMT. Both traces show a small decaying signal at early times due to light from the ablation plume and any ablation laser light that passes through the infrared filter. When the probe laser is on, the YbF LIF signal is observed as an 80 µs wide peak centred some 250 µs after the ablation pulse. For each trace, any constant background light is subtracted and then the trace obtained with the probe laser off is subtracted from the one with the probe laser on. The resulting background-free photon rates for both the signal and reference are integrated over the time window between 120 µs and 600 µs. The ratio yields a relative LIF signal that is independent of the molecular flux. The signals for each filter, together with the wavelength-dependent response of the detection system, determine the Franck-Condon factors which are presented in Table I. The quoted uncertainties are dominated by a 5% uncertainty in the relative PMT efficiency at different wavelengths, and to a lesser extent by residual variations in the measured signals due to the polarisation dependence of the filters.

A DLIF spectrum obtained at ASU from excitation of the $^{174}$YbF $R_1(2)$ branch feature is shown in Figure 3. The spectrum was obtained using a 0.1 nm step size for the scanning monochromator and averaging 40 ablation pulses per step. Spectra recorded under the same conditions but with the ablation laser blocked and separately with the excitation laser blocked were subtracted to remove very small chemiluminescence and background light from the excitation laser. DLIF spectra resulting from single-mode dye laser excitation of both the $R_1(2)$ branch feature and near the $P_1+Q_{12}$ band head ($N'' \approx 4$) were recorded multiple times. The integrated areas of the spectral features were used to determine the relative DLIF intensities from which the $f_{v'-v''}$ of Table I were extracted. The quoted errors represent the 1σ statistical uncertainties determined from multiple measurements. Systematic errors are estimated to be ≤ 1%. The values for $f_{v'-v''}$ obtained at the two excitation frequencies and in the two setups (IC and ASU) are consistent with one another, and we take their weighted means to obtain the final experimental values.

The decay curve resulting from pulsed dye laser excitation of the $P_1+Q_{12}$ band head ($N''\approx 4$) is shown in Figure 4. The fit to a single exponential curve, with the 10 ns pulse width of the laser excluded, gives a characteristic decay time of $28 \pm 2$ ns. The YbF molecules spend approximately 5 μs in the LIF collection zone, and since this is far longer than the measured decay time there is no distortion of the decay profile due to molecules exiting the detection region.

## Discussion

The suite of programs developed by Prof. Robert LeRoy (Waterloo University) [18] was used to predict the Franck-Condon factors. The potential energy curves were calculated using the first-order Rydberg-Klein-Rees (RKR1) procedure. The rotational parameters for the $A^2\Pi_{1/2}$ and $X^2\Sigma^+$ state were taken from Ref. 14 and the vibrational parameters from Ref. 7. The ro-vibronic wavefunctions were numerically evaluated to predict the Franck-Condon factors. The results are compared with experimental values in Table I. The measured value for $f_{0-0}$ is slightly larger than we calculate, and correspondingly the measured $f_{0-1}$ is slightly smaller than calculated.

The radiative lifetime of the $A(v'=0)$ state, $\tau$, is given by $\tau^{-1} = 3.137 \times 10^{-7} |\mu_{A-X}|^2 \sum_{v''} f_{0-v''} \nu_{0,v''}^3$, where the transition dipole moment, $\mu_{A-X}$, is in Debye (D), and the transition wavenumbers $\nu_{0,v''}$, are in cm$^{-1}$. From our measured $\tau$ and $f_{0-v''}$ we obtain $\mu_{A-X} = 4.39 \pm 0.16$ D, where the uncertainty is dominated by the uncertainty in the lifetime. The lifetime and transition moment are similar to those of CaF [19] and SrF [1].

Efficient laser cooling requires the molecule to fluoresce to a minimal number of ro-vibronic levels so that only a small number of laser frequencies are needed to keep it in the cooling cycle. Let us first consider the rotational branching. Neglecting small hyperfine splittings due to nuclear-spin interactions, each state of $A^2\Pi_{1/2}$ ($v=0, J, p$), specified by the total angular momentum $J$ and the parity $p$, decays to three levels within each vibrational state of $X^2\Sigma^+$ ($\Delta J=0, \pm 1$). The lowest two levels of $A^2\Pi_{1/2}$ are exceptions, because they have $J=1/2$ and so have only two decay routes ($\Delta J=0, +1$) to each vibrational state of $X$.

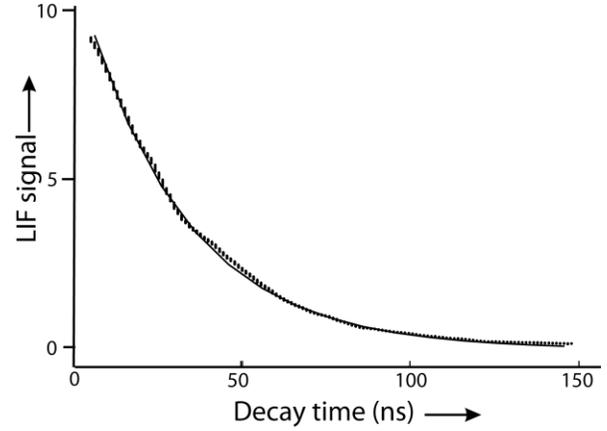

**Figure 4**: The fluorescence decay curve resulting from excitation of the $P_1+Q_{12}$ band head ($\nu=18106$ cm$^{-1}$).

For the $A^2\Pi_{1/2}$ ($v=0, J=1/2$) component of positive parity the decay channels are the $P_1(1)$ and $^PQ_{12}(1)$ branch features, which are separated only by the spin-rotation splitting of order 10 MHz. This is easily spanned by adding radio-frequency (rf) sidebands to the laser light. Each line is further split into two hyperfine components due to the fluorine nuclear spin, but these are also easily spanned using rf sidebands. The SrF laser cooling experiment [1,2] utilized these branch features. Unlike SrF, and most other molecules, the equilibrium bond distance for YbF is smaller in the $A^2\Pi_{1/2}$ state than in the $X^2\Sigma^+$ state. Consequently, low-$N$ band heads form in the $P_1$ and $^PQ_{12}$ branches in YbF and so the $P_1(1)$ and $^PQ_{12}(1)$ lines lie in a severely congested part of the spectrum. This is illustrated in Figure 5 where the observed pulsed-dye laser excitation spectrum of the $A^2\Pi_{1/2} \leftarrow X^2\Sigma^+$ (0,0) band of a supersonic YbF beam is compared with the predicted spectrum. Only the $^OP_{12}$ branch is free from overlap. The high density of overlapping lines does not pose any serious problem for laser cooling, but does makes it a little more difficult to locate the correct transitions.

For the component of $A^2\Pi_{1/2}$ ($v=0, J=1/2$) with negative parity the decay channels are the $Q_1(0)$ and $^OP_{12}(2)$ branch features. These features lie in less congested parts of the spectrum, particularly the $^OP_{12}(2)$ line where there is no overlap with any other lines. However, these two branch features are separated by $\approx 6B$ (43 GHz) which is a considerably less convenient frequency separation to span.

**Table I.** Measured and predicted Franck-Condon factors, $f_{v'-v''}$, for the $A^2\Pi_{1/2}(v'=0) \to X^2\Sigma^+(v'')$ transition of YbF.

| $f_{v'-v''}$ | IC$^a$ | ASU$^b$ | ASU$^c$ | Combined$^d$ | Predicted$^e$ |
|---|---|---|---|---|---|
| $f_{0-0}$ | 0.928±0.005 | 0.927±0.010 | 0.937±0.004 | 0.933±0.003 | 0.915 |
| $f_{0-1}$ | 0.069±0.005 | 0.073±0.010 | 0.063±0.004 | 0.066±0.003 | 0.083 |
| $f_{0-2}$ | (3±0.5)×10$^{-3}$ | <5×10$^{-3}$ | <5×10$^{-3}$ | (3±0.5)×10$^{-3}$ | 2.7×10$^{-3}$ |
| $\Sigma_{n>2} f_{0-n}$ | ($0^{+5}_{-0}$)×10$^{-4}$ | | | ($0^{+5}_{-0}$)×10$^{-4}$ | |

$^a$ Imperial College with $R(2)$ excitation. $^b$ Arizona State University with $R(2)$ excitation. $^c$ Arizona State University with $P_1+Q_{12}$ band head excitation. $^d$ Weighted mean of measurements. $^e$ $R(2)$ transition predicted using an RKR1 potential with the parameters (cm$^{-1}$): $B_e'' = 0.241294$, $B_e' = 0.247629$, $D_e'' = 2.388\times10^{-7}$, $D_e' = 1.999\times10^{-7}$, $\omega_e'' = 506.674$, $\omega_e x_e'' = 2.245$, $\omega_e' = 537.0$, $\omega_e x_e' = 3.0$.

With the use of three lasers it is possible to drive transitions from $v''=0$, 1 and 2 so that molecules remain in the cooling cycle until they decay from $v'=0$ to $v''>2$. Our measurements place an upper bound of $5\times10^{-4}$ on the probability of this unwanted decay, indicating that a YbF molecule can scatter at least 2000 photons on average before it decouples from the cooling light. This will allow substantial cooling of the translational motion of a supersonic or buffer-gas-cooled beam. Laser deceleration of the beam requires more photons to be scattered and may not be feasible, but the laser-cooled beam can be decelerated using conservative forces, e.g using a travelling-trap Stark decelerator [20, 21]. Together, Stark deceleration and laser cooling of a buffer-gas-cooled beam of YbF will allow ultra-cold slow-moving YbF pulses to be produced, with enormous potential for improving the measurement of the electron's electric dipole moment.

## Acknowledgements

The research at Arizona State University has been supported by the National Science Foundation (Grant No.0646473). The research at Imperial College has been supported by the EPSRC, the STFC and the Royal Society.

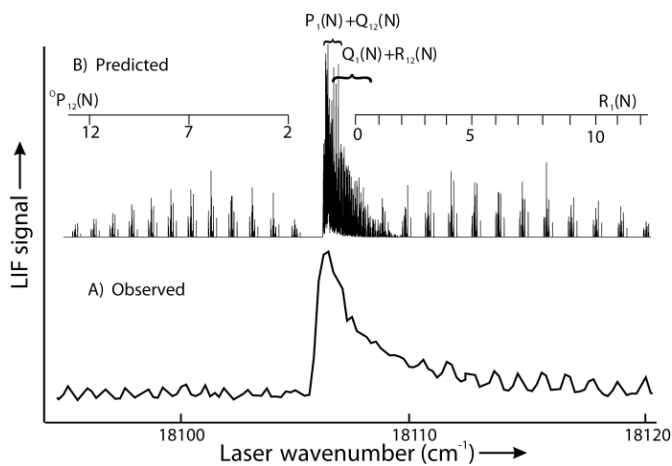

**Figure 5**: The observed (bottom) and predicted (top) broad-band LIF spectrum of the $A^2\Pi_{1/2} \leftarrow X^2\Sigma^+$ (0,0) band system of YbF. The observed spectrum was recorded using pulsed dye laser excitation with a 0.05 cm$^{-1}$ spectral resolution. The predicted spectrum was calculated using the spectroscopic parameters of Ref.13, a line width of 20 MHz and a rotational temperature of 20 K.